\documentclass[aps,prb,twocolumn,nofootinbib,floatfix]{revtex4-2}
\usepackage{amsmath,amssymb,graphicx,xcolor}
\usepackage{csquotes}

\begin{document}

\title{Not All Uniform B-Fields Are the Same!}
\author{Benjamin K. Luna$^1$}
\author{Murray S. Daw$^1$}
\affiliation{$^1$Clemson University}

\date{\today}


\begin{abstract}

Set up a charged rectangular-plate capacitor in a uniform $\mathbf{B}$-field. Quasi-statically turn off the $\mathbf{B}$-field – what happens? We show that this seemingly simple
induction problem (which appears in a famous undergraduate electrodynamics textbook) highlights the causal structure of electrodynamics. It may be (as it is in this problem) that different source configurations can produce the same uniform $\mathbf{B}$-field over some finite region, but very different dynamics when the source is turned off. We see also that the vector potential gives more information about the source of the $\mathbf{B}$-field, and is a very useful (and under-appreciated) way to understand induction and other key aspects of electrodynamics.

\end{abstract}

\maketitle

\section{Introduction}

In the recent history of classical electrodynamics, there has been a marked
trend towards the exclusion of the scalar and vector potentials in
favor of the electric $\mathbf{E}$- and magnetic
$\mathbf{B}$-fields, particularly as taught at the undergraduate level. However, the
originators of the theoretical formalism of electrodynamics
(F. E. Neumann, W. E. Weber, G. R. Kirchhoff, Lord Kelvin, and most
importantly J. C. Maxwell) made extensive use of the vector potential
$\mathbf{A}$. \cite{wu} This original usage hints at the little-known fact that the $\mathbf{A}$-field carries important information about the current source, which leads to a very natural usage of it in solving induction exercises. Following the physical insights in \textit{Physics for
Realists: Electricity and Magnetism} \cite{pfrem}, we will illustrate how this usage indeed facilitates solving such problems and gives greater insight into the causal structure of electrodynamics. Bodies with charge cause E\&M fields in the surrounding space, and those fields cause momentum in other charges. Fields by themselves make no sense without some (at least implicit) reference to the charges which cause them. 

\noindent \emph{Note: We use \emph{cgs} units.} \cite{jackson}

\section{A very simple but surprisingly insightful physics problem}

\begin{figure}[ht]
	\centering	\includegraphics[width=3in]{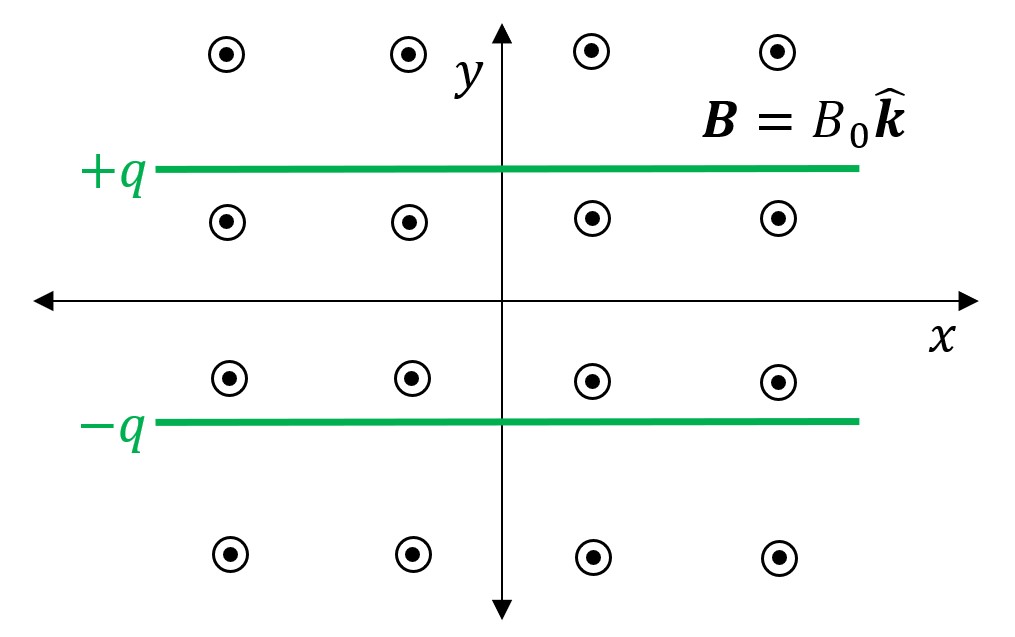}
	\caption{Edge-on view of a parallel-plate capacitor with square plates of side length $a$ and plate separation $d$ in a uniform $\mathbf{B}$-field, where the field lines are parallel to the planes in which the charged plates lie.}
	\label{InitialSetup}
\end{figure}

Consider a pair of square parallel plates with high mass and opposite charges \cite{insulator} $q$ and
$-q$ distributed uniformly on respective sides (where the charges are fixed in their positions on the plates) and square side length $a$ and plate
separation $d$, and place this capacitor at rest such that the field lines
of a uniform magnetic $\mathbf{B}$-field \cite{hint} in the $z$-direction
$\mathbf{B} = B_0 \hat{\mathbf{z}}$ are parallel to the planes in which those plates lie, shown in Fig. \ref{InitialSetup}. Quasistatically turn down the
$\mathbf{B}$-field to zero field strength - now, what are the forces
on the capacitor plates during the turnoff (these may not be uniform)
and what is the net impulse? 

\subsection{A common (but erroneous) method of solution} \label{commonMethod}

\begin{figure}[ht]
	\centering	\includegraphics[width=3in]{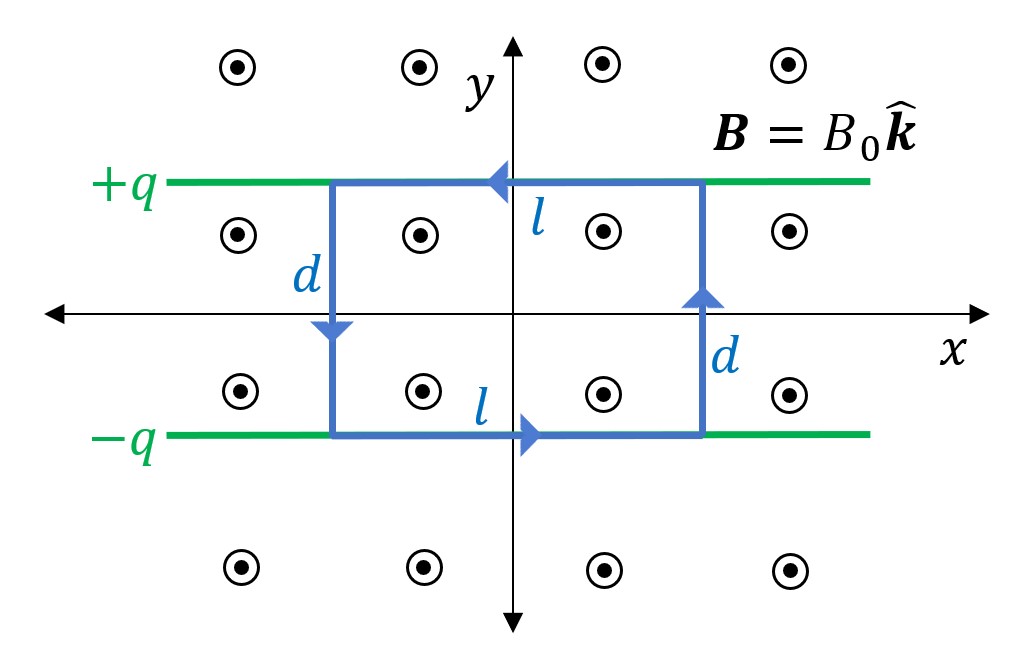}
	\caption{A rectangular Faraday closed path with two sides of length $l$ lying on the capacitor plates and two sides of length $d$ which cross the distance between the plates.}
	\label{FaradayLoop}
\end{figure}

The authors of this paper have posed this question to professional physicists and
students in an informal poll, and they have found that a common way of
approach starts with Faraday's law in the flux form. To illustrate one such approach, we take a
rectangular closed path inside the capacitor with the normal area vector
parallel to the field, one side $d$ in the $y$ direction crossing the region between the plates
and another side length $l$ in the $x$ direction lying on the plate, as shown in Fig. \ref{FaradayLoop}. Faraday's law for this path gives:
\begin{align}
\begin{split}
    \oint \mathbf{E} \cdot d \mathbf{l} &= - \frac{1}{c} \dot{\Phi} \\
    &= - \frac{1}{c} \dot{B} ld. \label{symmetryInOpp}
\end{split}
\end{align}
It is then implied (but not \textit{clearly} stated) that the only induced electric field contributing to the line integral is parallel to the planes of the
capacitor plates (after all, the only charge available to receive any impulse would be on the plates, not in between), and (also not clearly stated) we have the assumption that this induced electric field is constant over the plates, and so the legs of length $d$ going between the
capacitor are identified as contributing nothing to the line integral:
\begin{align}
    -l E(d) + l E(0) &= -\frac{1}{c}\dot{B}ld \label{jump} \\
    E(d) - E(0) &= \frac{1}{c}\dot{B}d. \label{faradayresult}
\end{align}
In fact, the step (\ref{jump}) is a jump - it does not follow from the
previous without an additional assumption. Nevertheless,
the force on the capacitor is identified from this to be (from the
Lorentz force) as
\begin{align*}
    \mathbf{F} &= -q E(0) \hat{\mathbf{i}} + q E(d) \hat{\mathbf{i}} \\
    &= q (E(d) - E(0)) \hat{\mathbf{i}} \\
    &= \frac{1}{c}q d \dot{B} \hat{\mathbf{i}},
\end{align*}
where Eq. (\ref{faradayresult}) was used in the last step. Now,
the impulse $\Delta \mathbf{p}$ is found as follows:
\begin{align*}
    \Delta \mathbf{p} &= \int_0^{\infty} \mathbf{F} dt \\
    &= \frac{1}{c}q d \hat{\mathbf{i}} \int_0^{\infty} \dot{B} dt \\
    &= \frac{1}{c}q d \hat{\mathbf{i}} (B(\infty) - B(0)) \\
    &= -\frac{q dB_0}{c} \hat{\mathbf{i}},
\end{align*}
which is the (false) answer desired in terms of the charge $q$. Remembering that the charge $q$ is related to the surface charge density $\sigma$ and the area of the plate $\mathcal{A}$ by
\begin{equation*}
	q = \sigma \mathcal{A}
\end{equation*}
and also that the electric field inside and far from the edges of the capacitor depends on the surface charge density by
\begin{equation*}
    E = 4 \pi \sigma,
\end{equation*}
we arrive at the (false) result desired in terms of the electric field between:
\begin{equation}
    \Delta \mathbf{p} = - \frac{EB_0 \mathcal{A} d}{4 \pi c} \hat{\mathbf{i}}. \label{falseResult}
\end{equation}


\noindent This is essentially the same as Problem 8.6 in Griffiths \cite{griffiths3rd} (3rd
edition only) in the chapter dealing with (among other things) the
Poynting vector. \cite{widespread}  In particular, the line of solution suggested in
that textbook was to calculate (approximately) the volume integral of
the Poynting Vector between the plates and to show that the impulse imparted to the plates while the
$\mathbf{B}$-field is turned off can be calculated using Eq. (\ref{falseResult}), and equals the total electromagnetic momentum (before turning off
$\mathbf{B}$) contained in the crossed $\mathbf{E}$- and $\mathbf{B}$-fields between the
plates. Others including McDonald \cite{mcdonald} and most especially Hu \cite{https://doi.org/10.48550/arxiv.1408.4144} have addressed this problem, which serves to illustrate the point that the source matters and must be sufficiently considered, the clear understanding of which is largely missing in the community of physicists. Following the publication of the third edition of the textbook mentioned above, Griffiths and
others noted in an AJP paper that the line of solution suggested by
the problem was ``almost entirely wrong''. \cite{babson} That paper showed
that the correct application of the Poynting Theorem involves
identifying a "hidden" momentum in the source of the $\mathbf{B}$-field (as
noted, ironically, in Griffiths' text just before Problem 8.6). In the
present paper, we show that, while of course correct in the sense of preserving momentum conservation, \cite{hiddenMom} this application
of the Poynting Theorem does not highlight the causal structure of
electromagnetism, which using the vector potential does very
nicely. In fact, this problem and a related one (which we discuss near
the end of this paper) clearly illustrate the causal role of the source by the use of the
vector potential $\mathbf{A}$. 

\subsection{What went wrong? A hidden assumption. (Well, in fact,
  there are two!)}

In the step (\ref{jump}) an extra assumption is made, as stated
previously. This assumption is usually not consciously made by those
who attempt this problem, which is why we must clearly state the steps
in the treatment. It is a misuse of Faraday's law.
 
Step (\ref{jump}) only applies if the source of the uniform $\mathbf{B}$-field
is one consisting of opposing planar currents. (As will be shown in
the following, it would \emph{not} be the correct answer, for
instance, if the uniform $\mathbf{B}$-field were established by a solenoid!)
This jump happens unconsciously to find what could be substituted into
Faraday's law. The whole approach to the solution does not hold
without this assumption about the source. In fact, this whole discussion brings to light a subtle point which we will expand upon later: no one who is unaware of the source knows how to solve the problem. In our pedagogical observations, the best one could do in ignorance of the source is this kind of guesswork.

Furthermore, as we shall detail in Section \ref{correctMethod}, there is a
\emph{second} assumption, which is that the capacitor plate is
arranged symmetrically inside of the source of the $\mathbf{B}$-field. This
second assumption is revealed by asking for the full force distribution
everywhere on the capacitor plates (not just the total impulse). 

\section{The correct approach: recognizing that the source matters} \label{correctMethod}

The simplest way to approach this exercise is to introduce the Lorentz force
law in terms of the scalar and vector potentials:
\begin{equation}
  \mathbf{F} = q (- \boldsymbol{\nabla} \phi - \frac{1}{c}
  \frac{\partial \mathbf{A}}{\partial t} + \mathbf{v} \times
  (\boldsymbol{\nabla} \times \mathbf{A})). \label{LorentzforceOG}
\end{equation}
(Although this is usually called the Lorentz force law, it is due to
Maxwell originally. \cite{fMaxwell, maxwell_2011_lorentzforce}) Now, we see that the dynamics of this exercise are contained
neither in the electrostatic term (for there is no $\phi$-field
generated apart from the capacitor) nor in the magnetostatic term (for
the capacitor is at rest and of high mass such that the capacitor does
not start to have an appreciable velocity until the field is
completely turned off). This leaves only the term in the middle, which
gives us an "induction $\mathbf{E}$" of
\begin{equation}
	\mathbf{E}_{\text{ind}} = - \frac{1}{c} \frac{\partial \mathbf{A}}{\partial t}.
	\label{secondPartMaxwellE}
\end{equation}
This will be remembered as the second part of "Maxwell's $\mathbf{E}$"
- in fact, this is also directly due to Maxwell. \cite{maxwell_2011} Upon integration with
respect to time, applying the initial and final conditions, and multiplying both sides by the charge, we see that
\begin{equation}
  \mathbf{p}_f = \frac{q}{c} \mathbf{A}_i, \label{potMomentum}
\end{equation}
where $\mathbf{p}_f$ indicates the final momentum and $\mathbf{A}_i$
indicates the initial vector potential, which shows that \textit{the vector
potential is a measure of \textbf{potential momentum per unit charge}} (or the maximum impulse per unit charge that the $\mathbf{A}$-field could impart to a massive charged body at that point), which
should be reminiscent of discussions in classical mechanics of the
canonical momentum of a charged point particle in an electromagnetic
field. This insight is also essentially due to Maxwell, which was
discussed more recently by Konopinski \cite{konopinskipaper}, Semon and Taylor \cite{semontaylor},
and Berche, Malterre, and Medina \cite{doi:10.1119/1.4955153}. Examples of modern textbooks
which revive this insight are Konopinski \cite{konopinski}, Zangwill \cite{zangwill}, Griffiths \cite{fGriffiths} \cite{griffiths4th},
and Rizzi \cite{pfrem}. The treatment in Rizzi properly grounds this insight by giving a much more fully physical explanation. Using the vector potential
reveals much more of the causal structure of
electromagnetism. It is important to note that this provides precisely the little bit of specification on the source necessary for the problem above. Although the source is not fully specified, this small amount of specification is enough for our purpose here. We demonstrate this below, following Rizzi. 

Another way of approach to the problem which respects the causal structure of electrodynamics is via the Jefimenko equations, which give the electric $\mathbf{E}-$ and magnetic $\mathbf{B}-$fields in terms of the sources: \cite{pfrem, jefimenkoEq}
\begin{align}
\begin{split}
	\mathbf{E} &= \int \frac{[\rho]}{R^2} \hat{\mathbf{R}} d^3 \mathbf{r}' + \int \frac{[\dot{\rho}]}{cR} \hat{\mathbf{R}} d^3 \mathbf{r}' \\ 
	& \quad \quad - \int \frac{[\partial \mathbf{J} / \partial t]}{c^2 R} d^3 \mathbf{r}' \\
\end{split}
\end{align}
\begin{align}
	\mathbf{B} &= \int \frac{[\mathbf{J}] \times \hat{\mathbf{R}}}{cR^2} d^3\mathbf{r}' + \int \frac{[\partial \mathbf{J} / \partial t] \times \hat{\mathbf{R}}}{c^2 R} d^3\mathbf{r}',
\end{align}
where $\mathbf{R} = \mathbf{r} - \mathbf{r}'$ is the vector pointing from the source element being integrated over to the field point, $R = \lvert \mathbf{R} \rvert$, $\hat{\mathbf{R}} = \mathbf{R}/R$, and the expressions in brackets are taken to be evaluated at the retarded time $t' = t - R/c$. One should compare the third term in the expression for $\mathbf{E}$ with Eq. (\ref{secondPartMaxwellE}) - for the Jefimenko's expression for $\mathbf{E}_\text{ind}$ in the quasistatic approximation (where the propagation time $R/c$ is negligible)
\begin{align*}
	\mathbf{E}_\text{ind} &= - \frac{1}{c} \frac{\partial}{\partial t} \Bigg( \frac{1}{c} \int \frac{\mathbf{J}}{R} d^3\mathbf{r}' \Bigg) \\
	\mathbf{E}_\text{ind} &= - \frac{1}{c} \frac{\partial \mathbf{A}}{\partial t}
\end{align*}
which is the same as Eq. (\ref{secondPartMaxwellE}). We have here noted that (again in the quasistatic approximation)
\begin{align*}
	\mathbf{A} = \frac{1}{c} \int \frac{\mathbf{J}}{R} d^3\mathbf{r}'.
\end{align*}
Here, we see that the vector potential $\mathbf{A}$ affords a natural means to capture what is necessary about the source for induction, and, as mentioned above, it can be interpreted as a potential momentum per unit charge (in a manner analogous to how the scalar potential can be interpreted as a potential energy per unit charge). This way incorporates the same specification about the source as the first way, because it is using the $\mathbf{A}$-field without acknowledging it.

Another way to is to specify boundary conditions for the fields. This incorporates the same kind of information about the source as the above two ways, but without the clarity of insight of the first way.


\subsection{Equations expressing causal relations}

Using the Lorenz \cite{fLorenzName} gauge condition
\begin{equation}
	\boldsymbol{\nabla} \cdot \mathbf{A} = - \frac{1}{c} \frac{\partial \phi}{\partial t} \label{lorenzOG}
\end{equation}
in the Maxwell's equations for the divergence of $\mathbf{E}$ and the curl of $\mathbf{B}$, we have
\begin{align*}
	\boldsymbol{\nabla} \cdot \mathbf{E} &= 4 \pi \rho \\
	- \boldsymbol{\nabla} \cdot \bigg( \frac{1}{c} \frac{\partial \mathbf{A}}{\partial t} + \boldsymbol{\nabla} \phi \bigg) &= 4 \pi \rho \\
	- \frac{1}{c} \frac{\partial}{\partial t} \bigg( \boldsymbol{\nabla} \cdot \mathbf{A} \bigg) - \boldsymbol{\nabla}^2 \phi &= 4 \pi \rho \\
	\frac{1}{c^2} \frac{\partial^2 \mathbf{\phi}}{\partial t^2} - \boldsymbol{\nabla}^2 \phi &= 4 \pi \rho
\end{align*}
and
\begin{align*}
	\boldsymbol{\nabla} \times \mathbf{B} &= \frac{4 \pi \mathbf{J}}{c} + \frac{1}{c} \frac{\partial \mathbf{E}}{\partial t} \\
	\boldsymbol{\nabla} \times (\boldsymbol{\nabla} \times \mathbf{A}) &= \frac{4 \pi \mathbf{J}}{c} - \frac{1}{c} \frac{\partial \boldsymbol{\nabla} \phi}{\partial t} - \frac{1}{c^2} \frac{\partial^2 \mathbf{A}}{\partial t^2} \\
	\boldsymbol{\nabla} (\boldsymbol{\nabla} \cdot \mathbf{A}) - \boldsymbol{\nabla}^2 \mathbf{A} &= \frac{4 \pi \mathbf{J}}{c} - \boldsymbol{\nabla} \bigg( \frac{1}{c} \frac{\partial \phi}{\partial t} \bigg) - \frac{1}{c^2} \frac{\partial^2 \mathbf{A}}{\partial t^2} \\
	\frac{1}{c^2} \frac{\partial^2 \mathbf{A}}{\partial t^2} - \boldsymbol{\nabla}^2 \mathbf{A} &= - \boldsymbol{\nabla} \bigg( \boldsymbol{\nabla} \cdot \mathbf{A} + \frac{1}{c} \frac{\partial \phi}{\partial t} \bigg) + \frac{4 \pi \mathbf{J}}{c} \\
	\frac{1}{c^2} \frac{\partial^2 \mathbf{A}}{\partial t^2} - \boldsymbol{\nabla}^2 \mathbf{A} &= \frac{4 \pi \mathbf{J}}{c}.
\end{align*}
These are the inhomogeneous wave equations for the scalar and vector potentials, which allow expressions of the $\phi$- and $\mathbf{A}$-fields in terms of the charge density $\rho(\mathbf{r}, t)$ and current density $\mathbf{J}(\mathbf{r},t)$ that generate them. The solutions to these equations and the specifications we have stated above are called the retarded potentials, and these are given below \cite{fLocalSol}:
\begin{align}
\begin{split}
	\phi(\mathbf{r},t) &= \int \frac{[\rho]}{R} d^3 \mathbf{r}'	\\
	\mathbf{A}(\mathbf{r},t) &= \frac{1}{c} \int \frac{[\mathbf{J}]}{R} d^3 \mathbf{r}'. 
\end{split} \label{vecPotRetarded}
\end{align}
This form makes clear that $\phi(\mathbf{r}, t)$ is determined by the distribution of the charge density in the source, and $\mathbf{A}(\mathbf{r}, t)$ is determined by the distribution of current density in the source, taking into account the propagation time $R/c$. These retarded potentials will yield the Jefimenko equations mentioned in the previous subsection.

For the purposes of this paper, the above retarded potentials may be simplified, since we deal with the quasistatic case, which is where the current is turned off slowly enough so that the time of propagation can be neglected. This simplification consists in recognizing that the retarded time will only be negligibly different from $t$, and so we may replace the retarded time in the above potential expressions with $t$ (which reduces our expressions to the quasistatic ones):
\begin{align}
\begin{split}
	\phi(\mathbf{r},t) &= \int \frac{\rho(\mathbf{r}', t)}{\lvert \mathbf{r} - \mathbf{r}' \rvert} d^3 \mathbf{r}'	\\
	\mathbf{A}(\mathbf{r},t) &= \frac{1}{c} \int \frac{\mathbf{J}(\mathbf{r}', t)}{\lvert \mathbf{r} - \mathbf{r}' \rvert} d^3 \mathbf{r}'. 
\end{split} \label{vecPotQS}
\end{align}
Although we derived these expressions in the Lorenz gauge, we see that these forms for the retarded quasistatic potentials in the Lorenz gauge are identical in form to the quasistatic Coulomb gauge forms. \cite{fLorenzCoulomb} Without considering explicitly every detail of propagation, these forms clearly show the causal relations between the sources of charge and current ($\rho$ and $\mathbf{J}$) and the fields ($\phi$ and $\mathbf{A}$).




\subsection{One choice of source and positioning the capacitor within}

Instead of \textit{unconsciously} choosing a source like the above
common method detailed in Section \ref{commonMethod}, we may \textit{consciously} choose the source of
infinite opposing planar currents with current magnitude per
perpendicular length $\mathcal{I} = \frac{cB_0}{4 \pi}$, 
\begin{figure}[ht]
	\centering
	\includegraphics[width = 3in]{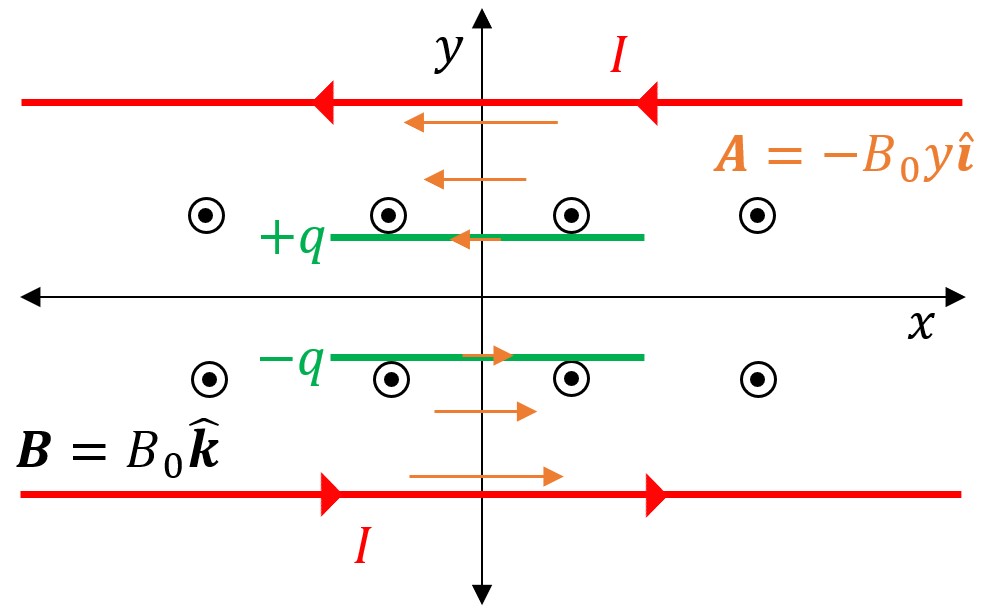}
	\caption{Infinite opposing planar currents, with $\mathbf{B}$
          inside shown on the left and $\mathbf{A}$ inside shown on the right.} \label{OpposingPlanarSource} 
\end{figure}
but we cannot forget that this is an assumption we have introduced -
this does not hold generally for all sources. The vector potential, obtained from Eq. (\ref{vecPotQS}), for
such a source is (choosing the origin of the coordinate system to be
centered within the source) 
\begin{equation}
	\mathbf{A} = -B_0 y \hat{\mathbf{i}},
\end{equation}
which may be quickly verified to produce the necessary
$\mathbf{B}$-field. \cite{fZeroth} At this point, we have only introduced a source
without any induction - the interesting physics is in the
induction. We introduce it by the following quasistatic prescription: 
\begin{equation}
	\mathcal{I}(t) = \frac{cB_0}{4 \pi} \bigg( 1 - \frac{t}{T}
        \bigg), \quad 0 \leq t \leq T. 
\end{equation}
Once we have this prescription, we see that putting this into the
vector potential (within the quasistatic approximation of the original
problem) and differentiating to find the induction $\mathbf{E}$ gives us 
\begin{equation}
	\mathbf{E}_{\text{ind}} = - \frac{B_0}{cT} y \hat{\mathbf{i}}.
\end{equation}
Now, we see that another assumption comes into play - we place the
capacitor with its plates not only parallel to the $\mathbf{B}$-field
(as specified in the original problem) but parallel to the planar
currents as well, with the positively charged plate on the positive
side of the $y$-axis and centered on the midplane $y = 0$. Without this assumption, it is not necessary that the $\mathbf{E}$-field is constant over the extension of the plates. Integrating
the force given by the induction $\mathbf{E}$ over the capacitor, we
obtain the total impulse 
\begin{equation}
	\Delta \mathbf{p} = -\frac{qdB_0}{c} \hat{\mathbf{i}},
\end{equation}
and noting that both plates receive equal halves of this impulse, such
that zero total angular momentum is imparted with respect to the
center of mass of the capacitor (located on the midplane here). 

\begin{figure}[ht]
	\centering
	\includegraphics[width=3in]{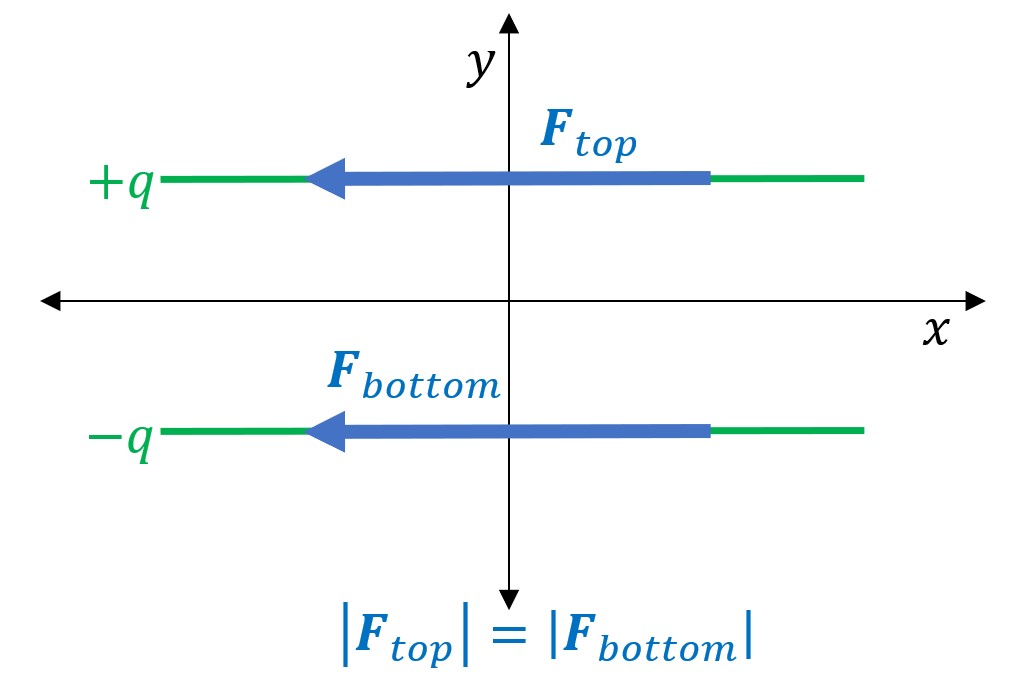}
	\caption{The arrows shown are the net forces on the
          capacitor during the ramp-down of the magnetic field in
          Fig. \ref{OpposingPlanarSource} - the net forces on the top
          plate and the bottom plate are equal.} \label{optionA} 
\end{figure}
So far, we have the same result as before, shown in Fig. \ref{optionA}. 

\subsection{A different position for the capacitor plates}

\begin{figure}[ht]
	\centering
	\includegraphics[width = 3in]{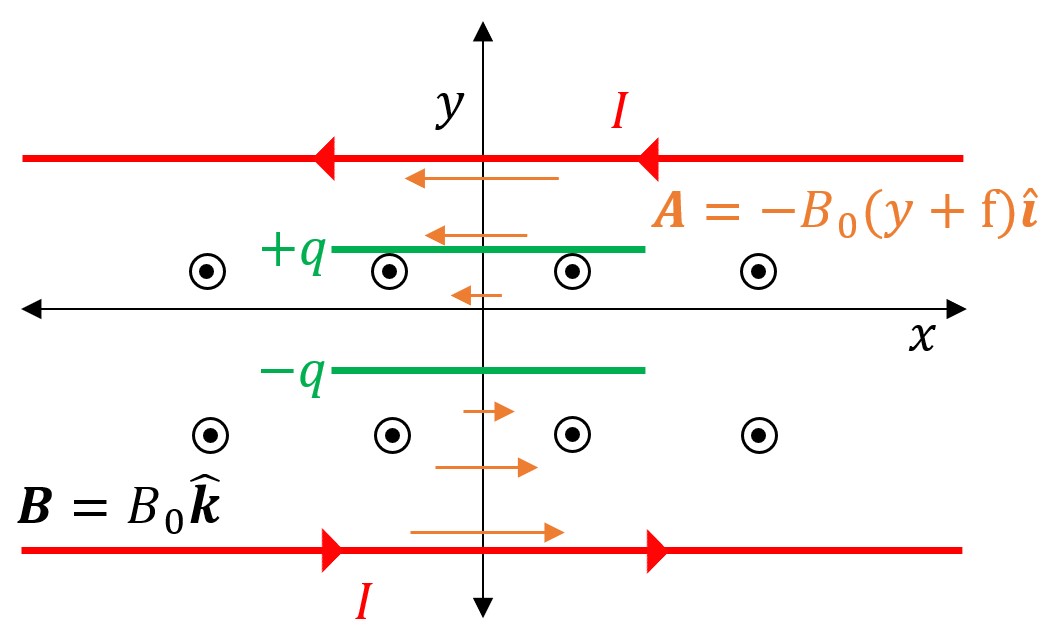}
	\caption{Infinite opposing planar currents shifted down by a distance $f$.}
	\label{oppPlanarShiftedTemp}
\end{figure}

We noted two assumptions made in the previous section - the choice of
source current and the  choice of positioning of the capacitor
relative to that current. To see whether these are arbitrary so as not
to change the dynamics if they are changed, we first change the
capacitor's position while maintaining the same source. Instead of
having the capacitor centered on the midplane between the opposing
planar currents, we shift the source down (or, equivalently, the
capacitor up) by some distance $f$ while maintaining its
plate-parallel orientation with respect to the planar currents, as shown in Fig. \ref{oppPlanarShiftedTemp}. 

\noindent Note that $\mathbf{B}$ is exactly the same as before. We turn off the
current by the same prescription as above. When we find the induction
$\mathbf{E}$, we find that the total impulse imparted to the capacitor
is the same as above:  
\begin{equation}
	\Delta \mathbf{p} = - \frac{qdB_0}{c} \hat{\mathbf{i}}.
\end{equation}
However, even though $\mathbf{B}$ is exactly the same as before everywhere on the capacitor, the
impulse is not divided evenly between the plates of the capacitor -
the top plate receives a greater impulse than the bottom, showing that
there is a net angular momentum about the center of mass of the
capacitor of  
\begin{equation}
	\Delta \mathbf{L} = \frac{qdfB_0}{c} \hat{\mathbf{k}},
\end{equation}
\begin{figure}[ht]
	\centering
	\includegraphics[width=3in]{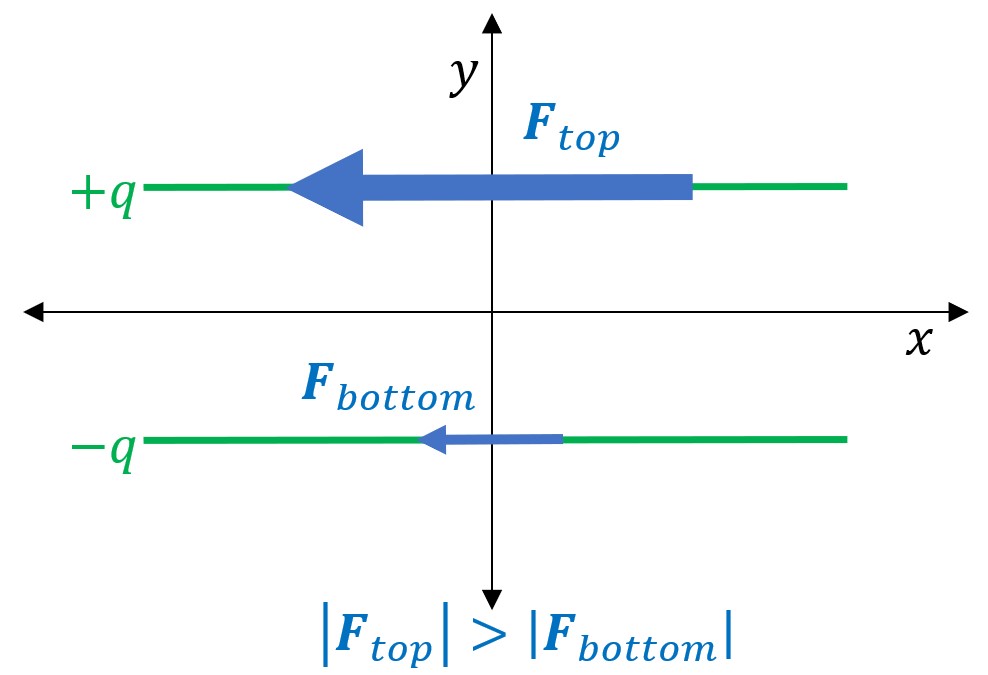}
	\caption{The arrows shown are the net forces on the
          capacitor during the ramp-down of the magnetic field in
          Fig. \ref{oppPlanarShiftedTemp} - note here the greater
          force on the top plate than on the bottom plate, which gives
          the whole capacitor a nonzero angular momentum.} 
	\label{optionBoriginal} 
\end{figure}
\noindent and this is shown in Fig. \ref{optionBoriginal}. 



\noindent We see from our solution that even though we have the same
$\mathbf{B}$-field on the capacitor, and even though we get the same net impulse
imparted to a capacitor in this field, the result is completely
different, as the capacitor in the first field receives a net zero
angular momentum and the capacitor in the second field receives a
nonzero angular momentum! We see that \textbf{the $\mathbf{A}$-field at the location of the capacitor is
  very obviously carrying some physical information about the source not found in the 
  $\mathbf{B}$-field there}. 

\subsection{A different orientation}

Perhaps not surprisingly, the story of the original Faraday's law
argument takes a turn for the worse. For now, we continue to leave the
source assumption (1) unaltered, and we keep the capacitor first
centered on the midplane as before, but we rotate the whole source either clockwise or
counter-clockwise about the $z$-axis by $90^{\circ}$.   
\begin{figure}
	\centering
	\includegraphics[width = 3in]{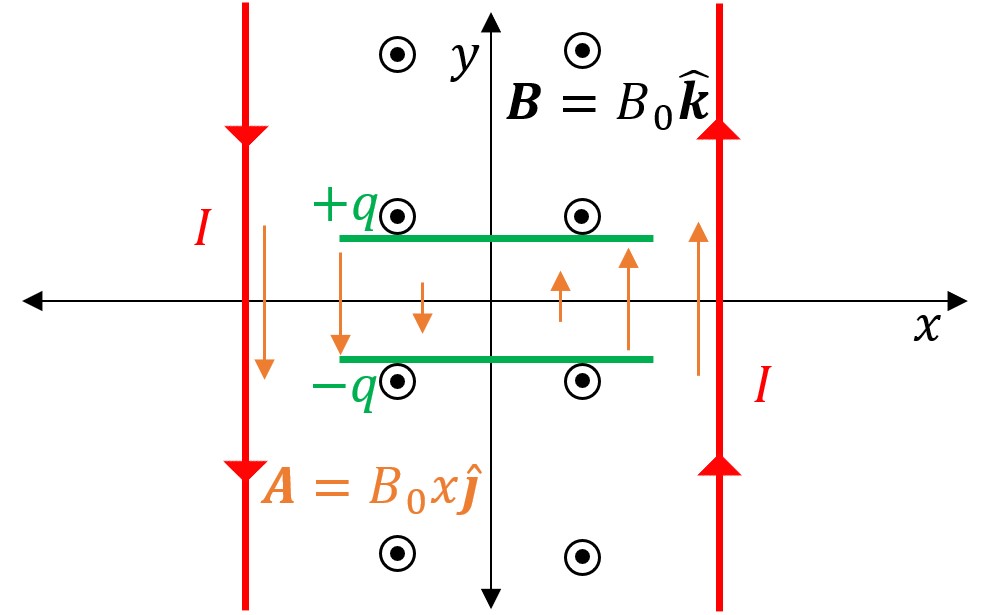}
	\caption{Infinite opposing planar currents, but rotated
          counter-clockwise about the $z$-axis by $90^{\circ}$.}  
	\label{rotatedOpposing}
\end{figure}
      
\begin{figure}[ht]
	\centering
	\includegraphics[width=3in]{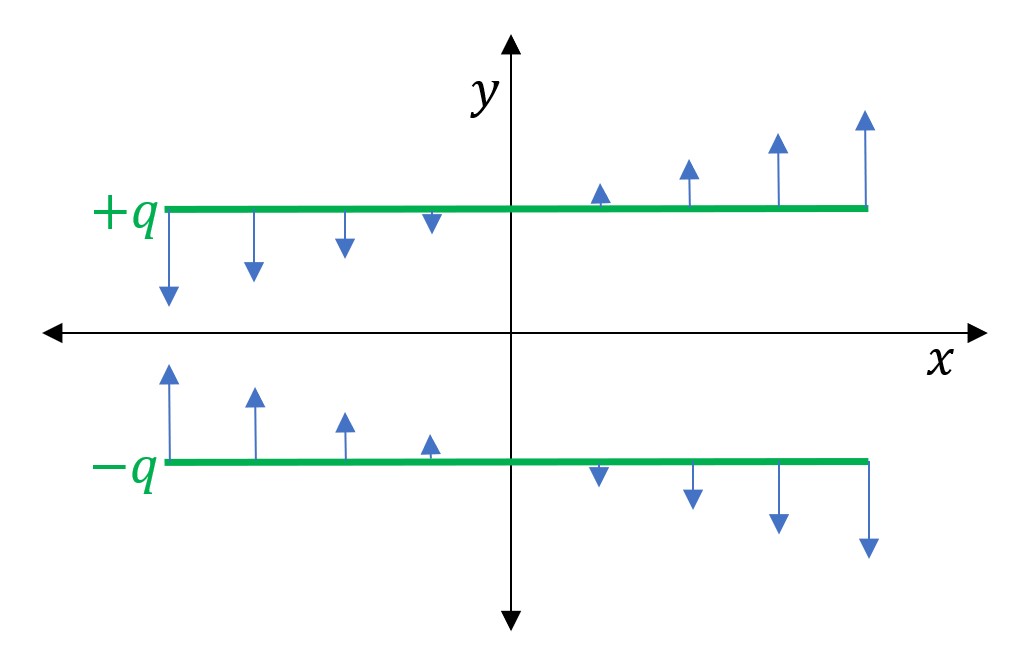}
	\caption{The arrows shown are the forces on the various parts of the
          capacitor during the ramp-down of the magnetic field in
          Fig. \ref{rotatedOpposing} - there is no net impulse to
          either plate, but there is a net torque on each plate.} 
	\label{optionB}
\end{figure}

This is shown in Fig. \ref{rotatedOpposing}. Note again that we have
precisely the same magnitude and direction of the
$\mathbf{B}$-field. The corresponding $\mathbf{A}$-field is \cite{GaugeQuestion} 
\begin{equation}
	\mathbf{A} = B_0 x\hat{\mathbf{j}}. 
\end{equation}
We execute the same process and note that the induced $\mathbf{E}$ is
no longer in the $\hat{\mathbf{i}}$ direction - we get for the total
impulse of the whole capacitor 
\begin{equation}
	\Delta \mathbf{p} = 0.
\end{equation}
Here, something else interesting happens - the individual plates are
each given an opposite angular momentum such that 
\begin{equation}
	\Delta \mathbf{L} = 0
\end{equation}
for the whole capacitor about the center of mass, but for the top
plate alone about its own center of mass (which is located on the
plate itself) 
\begin{equation}
	\Delta \mathbf{L}_{\text{top}} = \frac{qa^2 B_0}{12c} \hat{\mathbf{k}},
\end{equation}
and for the bottom plate alone about its own center of mass
\begin{equation}
	\Delta \mathbf{L}_{\text{bottom}} = - \frac{qa^2 B_0}{12c} \hat{\mathbf{k}},
\end{equation}
as shown in Fig. \ref{optionB}.

\subsection{Yet another source}

\begin{figure}[ht]
	\centering
	\includegraphics[width=3in]{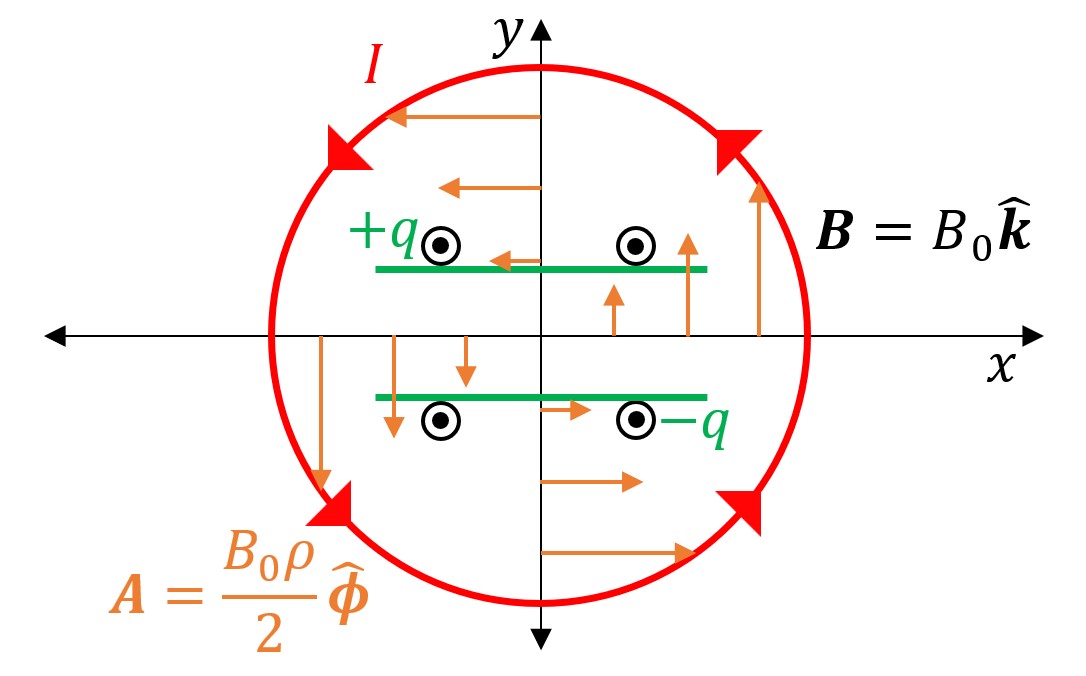}
	\caption{A solenoid of infinite length with $\mathbf{B}$-field pointing
          parallel to the axis.}  
	\label{solenoidSource}
\end{figure}

\begin{figure}[ht]
	\centering
	\includegraphics[width = 3in]{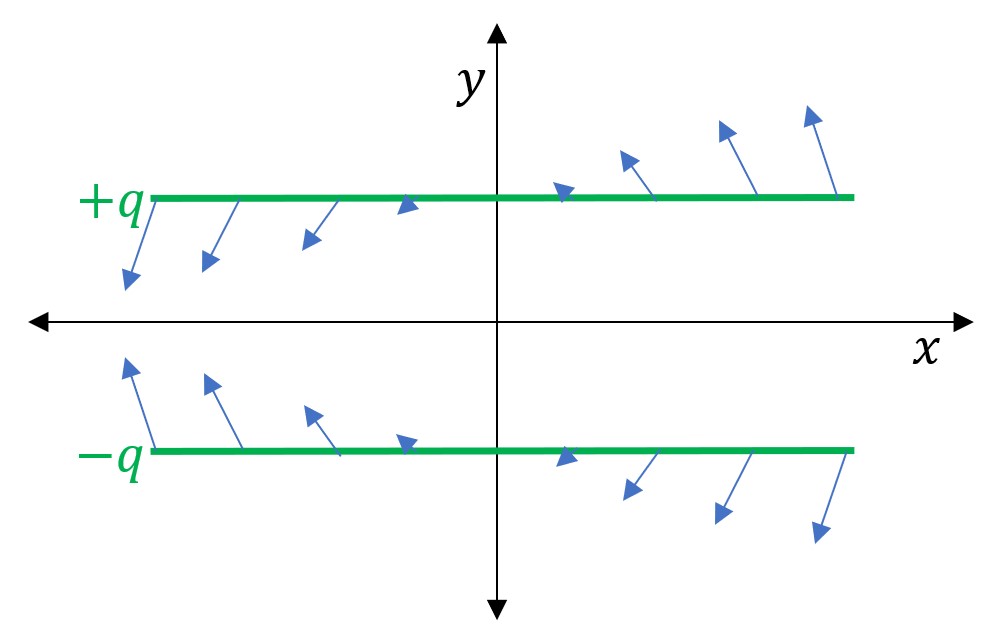}
	\caption{The arrows shown are the forces on the parts of the
          capacitor during the ramp-down of the magnetic field in
          Fig. \ref{solenoidSource} - there is a net impulse to both
          plates and a net torque on both plates as well.} 
          \label{optC}
\end{figure}

As a final example of how badly the original argument fails, we
finally consider a third case with a completely different kind of source -- a solenoid, as in
Fig. \ref{solenoidSource}.  
\noindent We have not changed the magnitude or the direction of the
$\mathbf{B}$-field, and the corresponding $\mathbf{A}$-field is 
\begin{equation}
	\mathbf{A} = \frac{B_0 \rho}{2} \hat{\boldsymbol{\phi}}
\end{equation}
(in cylindrical polar coordinates). We once again turn off the
planar current by the same prescription as before. This time, we have
a combination of the two effects we saw above. The total impulse
imparted to the whole capacitor is 
\begin{equation}
	\Delta \mathbf{p} = - \frac{qdB_0}{2c} \hat{\mathbf{i}},
\end{equation}
which is precisely half of the total impulse of the first two
situations we examined. The impulse is divided evenly between the
plates. We also see that although the total angular momentum of the
whole capacitor about its center of mass is zero, the angular momentum
of the top plate about its own center of mass is 
\begin{equation}
	\Delta \mathbf{L}_{\text{top}} = \frac{qa^2 B_0}{24c} \hat{\mathbf{k}}
\end{equation}
and that the angular momentum of the bottom plate about its own center of mass is
\begin{equation}
	\Delta \mathbf{L}_{\text{bottom}} = - \frac{qa^2 B_0}{24c} \hat{\mathbf{k}},
\end{equation}
as shown in Fig. \ref{optC}.

\section{A second exercise that should be simpler but which causes
  more concern for most}

\begin{figure}[ht]
	\centering
	\includegraphics[width=3in]{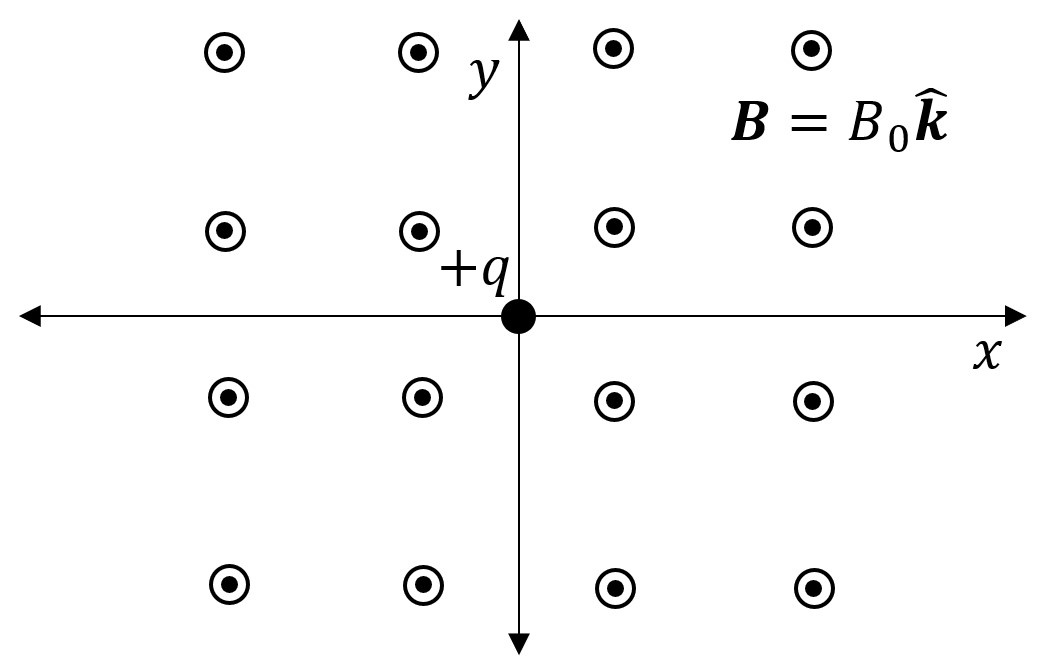}
	\caption{The arrows shown are the initial uniform $\mathbf{B}$-field with a point charge $+q$ in that region.}
	\label{ptChg}
\end{figure}

To bring the point home, we now slightly alter the original question
posed - instead of a capacitor, we place a point charge
in the field, as shown in Fig. \ref{ptChg} and turn off the $\mathbf{B}$-field. What happens? This problem captures the essence of Problem 7.21 in the fourth edition of Griffiths. \cite{griffiths4th} In our informal poll, the participants
found this problem to be even more puzzling than the first problem
with the capacitor, because they could find no ``handle'' -- there's obviously
no clear way to use Faraday's Law involving a path integral. 

The solution to this problem will involve the use of Eq. (\ref{potMomentum}), which we reproduce below:
\begin{equation}
	\mathbf{p}_f = \frac{q}{c} \mathbf{A}_i,
\end{equation}
and, as we continually stress, knowledge of the source is required in order to solve the problem, as with the above capacitor problem. It cannot be solved without some specification of the source, whether implicit or explicit. We see that the vector potential provides a very simple way to incorporate such consideration of the source. \cite{fSourceJefimenko}

\section{Concluding remarks}

We began by looking at a common but erroneous method of solving this
problem, and we systematically showed that the implicit assumptions
made therein cannot be justified on the grounds of the original problem as stated. In point of fact, this question of
what happens to the capacitor in a uniform $\mathbf{B}$-field ramped
down from full strength to zero \emph{cannot be answered} because it
does not sufficiently specify the physical situation! It is necessary
to specify the source of the $\mathbf{B}$-field in some way in order for the
problem to be well-posed, since bodies with charge are causally responsible for the fields in space, and these fields then impart momentum to other charges - this fact cannot be totally ignored. The confusion resulting from not being aware of this fact results in the guesswork outlined above. The vector potential provides a very simple and natural way to incorporate this information about the source. \cite{fADirection} Once we see this, we are able to correctly reformulate the question to be well-posed with the (necessary) specification of source: \textit{What happens to the the capacitor in a ramped-down uniform $B$-field caused by a particular source?}  

We clearly see that the $\mathbf{A}$-field carries information about the source not found in the (local) $\mathbf{B}$-field alone, and this is borne out in its elegant usage in solving induction problems. Different sources have different effects, even if that effect is not reflected in the $\mathbf{B}$-field at the charge's location. Not all uniform $\mathbf{B}$-fields are the same, and the consideration of the source is necessary in electrodynamics, since the fields make no sense without some (at least implicit) reference to the charged bodies which cause them in space. 

\section{Acknowledgments}

We are very grateful to Dr. Anthony Rizzi for extensive conversations and to Dr. Andrew Zangwill for helpful feedback.


\begin{thebibliography}{9}
\bibitem{insulator}
    Though the textbook source \cite{griffiths3rd} for this problem does state it for a 'capacitor', we have found it cleaner to demonstrate the principles involved using two uniformly charged insulator plates. The case of an insulator does not bring up the possible issue of mobile charges moving inside the plates, which would needlessly complicate the problem and make it more difficult to explain the principles involved. So the 'capacitor' in this paper is taken to represent two uniformly charged insulating plates.
\bibitem{hint}
      This vagueness should be disturbing to the reader. Specifically, one would like to know over what region the uniform $\mathbf{B}$-field is being specified. To have a unique answer to the question, we must specify some finite region within which the $\mathbf{B}$-field is arbitrarily uniform. We delay this discussion for reasons that will become apparent shortly.
\bibitem{widespread}
      It should be noted that this textbook is excellent in training students in the equations of electromagnetism, which is the reason for its just fame - a problem from this particular book is only singled out to illustrate the generic point about all physicists' understanding of electromagnetism broadly and would apply even more so to other authors of similar textbooks. The issues raised in this paper regarding the exercise above are not extraordinary - they are typical of the understanding of physicists and physics students.
\bibitem{hiddenMom}
    It was clarified by Hu in the above manuscript that finding the hidden momentum of the source of the $\mathbf{B}$-field is not necessary to obtain the impulse on the capacitor.
\bibitem{fLocalSol}
    These solutions of the inhomogeneous wave equations require that all of the sources are localized within some finite volume (that is, there are no sources $\rho$ or $\mathbf{J}$ at infinity).
\bibitem{fLorenzCoulomb}
    It should be noted that since the scalar potential in the cases that we will consider does not change, the Lorenz gauge condition, Eq. (\ref{lorenzOG}), reduces to 
    \begin{equation*}
	\boldsymbol{\nabla} \cdot \mathbf{A} = 0,
    \end{equation*}
    which is, interestingly, identical in form with the Coulomb gauge condition.
\bibitem{fZeroth}
    Note that this is a zeroth-order approximation to a set of finite sheet currents in $1/L$, where $L$ is half the length of each of the sides of the plates.
\bibitem{GaugeQuestion}
    Do not be deceived - this and the other changes in source mentioned do NOT constitute a difference of gauge. Gauge differences do not and cannot imply differences in dynamics.
\bibitem{fSourceJefimenko}
    Even if one does not wish to use the vector potential, one cannot avoid the (at least implicit) consideration of the source, as one clearly sees in the expression for $\mathbf{E}$ in the Jefimenko equations.
\bibitem{fADirection}
    In general, $\mathbf{A}$ will point in the direction of the current source (superposition), as is stated on page 248 of the fourth edition of Griffiths \cite{griffiths4th}, and as can easily be sketched based on Eq. (\ref{vecPotRetarded}) (and Eq. (\ref{vecPotQS}) in the quasistatic condition).
\end{thebibliography}
\end{document}